\documentclass[a4paper,11pt]{article}
\usepackage{jheppub}

\pdfoutput=1
\usepackage[T1]{fontenc}
\usepackage{CJK}
\usepackage{xcolor}
\usepackage{amsfonts}
\usepackage{epstopdf}
\usepackage{wrapfig} % to wrap figure with
\usepackage{subfigure}
\usepackage{graphicx}  % Include figure files
\usepackage{dcolumn}   % Align table columns
\usepackage{bm}
\usepackage{float}
\newcommand{\be}{\begin{equation}}
\newcommand{\ee}{\end{equation}}
 \newcommand{\bea}{\begin{eqnarray}}
 \newcommand{\ena}{\end{eqnarray}}

\title{Thermodynamics and weak cosmic censorship conjecture in Born-Infeld-anti-de Sitter black holes}

\author[1,2]{Xiao-Xiong Zeng,}
\author[3]{Hai-Qing Zhang}

\affiliation[1]{State Key Laboratory of Mountain Bridge and Tunnel Engineering, Chongqing Jiaotong  University, Chongqing 400074, China}
\affiliation[2]{Department of Mechanics, Chongqing Jiaotong University, Chongqing 400074, China}
\affiliation[3] {{Center for Gravitational Physics,} Department of Space Science and International Research Institute of Multidisciplinary Science, Beihang University, Beijing 100191, China}

\emailAdd{xxzengphysics@163.com}
\emailAdd{hqzhang@buaa.edu.cn}

\abstract{\\   We examine the laws of thermodynamics and the weak cosmic censorship conjecture in the normal and extended phase space of Born-Infeld-anti-de Sitter black hole by considering a charged particle absorption. In the normal phase space, the first and second law of thermodynamics as well as the weak cosmic censorship are still valid. However, in the extended phase space, the second law of thermodynamics is violated for double-horizon black holes and part of single-horizon black holes. The first law of thermodynamics and the weak cosmic censorship conjecture are still valid for all black holes. In addition, we find that the shift of the metric function,  which determines the locations of the horizons, takes the same form at the minimum point in both the normal and extended phase space, indicating that the weak cosmic censorship conjecture is independent of the thermodynamic phase space.   }

\begin{document}
\maketitle
%\flushbottom

\section{Introduction}

Since the pioneering work of Bekenstein \cite{Bekenstein:1973ur,Bekenstein:1974ax} and Hawking \cite{Hawking:1974sw,Hawking:1976de}, black hole thermodynamics has become a fundamental topic in general relativity. Temperature and entropy in the normal thermodynamical systems are  related to the horizon of a black hole. Until now, there have been numerous methods to derive the temperature and entropy of a black hole \cite{Hartle,Gibbons,Damour,Parikh,Iso,tHooft:1984kcu,Cardy:1986ie}.  It was found that the temperature $T$ of the black hole was related to the surface gravity $\kappa$ at the horizon, i.e., $T=\kappa/(2 \pi)$, and the entropy was proportional to the area of the horizon.  The horizon is important not only for the thermodynamics, but also for the causality of the spacetime. If there is no horizon, the singularity will be naked for an infinity observer which makes the causality break down. But due to Penrose\cite{Penrose:1969pc}, all singularities arising from gravitational collapse must be hidden by the black hole horizons, which is the so-called ``weak cosmic censorship conjecture''.

The laws of thermodynamics and weak cosmic censorship conjecture can be tested   from a charged particle   absorption by a black hole.
As a particle drops into the black hole, it has been shown that the first law and   the   second law of black hole thermodynamics   still hold \cite{Gwak:2015fsa,Gwak:2016gwj}. While for the weak cosmic censorship conjecture, there are still some debates.  Wald proposed firstly a gedanken experiment to check this conjecture in the extremal  Kerr-Newman black hole, {which showed} that no violations occur  as a particle is   thrown
into a black hole \cite{Wald:1974ge}. Nevertheless,  the conjecture  would be violated in the   near-extremal Reissner-Nordstrom \cite{Hubeny:1998ga} black hole  and near-extremal Kerr black hole \cite{Jacobson:2009kt}.
Later it was found that as the backreaction and self-force effects are taken into account, particles may be escaped from black holes and naked  singularities can be avoided \cite{Barausse:2010ka}.
  There have been many works discussing about the validity of  weak cosmic censorship conjecture in various black hole backgrounds under charged or spinning particles
    absorption \cite{Colleoni:2015afa,BouhmadiLopez:2010vc,Hod:2016hqx,Natario:2016bay,
  Horowitz:2016ezu,Duztas:2016xfg,Gwak:2017icn,Gao:2012ca,Rocha:2014jma,Rocha:2011wp,Chen:2018yah}. { Now we know that  including} the self-force effect or backreaction, both the extremal black holes and near-extremal black holes cannot be overcharged.
  This result was confirmed again   by considering the second-order corrections to mass  recently \cite{Wald2018,Sorce:2017dst,Ge:2017vun,An:2017phb}. It should be stressed that there is also a counter example to the weak cosmic censorship conjecture  in four-dimensional anti de Sitter(AdS)  space-time \cite{Crisford2017}, in which the curvature grows without bound in the future, leaving regions of spacetime with arbitrarily large curvatures naked to the infinite boundary observers.

Thermodynamics in  AdS space now prevails. One possible reason is the application of the AdS/CFT correspondence \cite{Maldacena:1997re,Gubser:1998bc,Witten:1998qj,Aharony:1999ti}, which
relates the gravity theory in $D$-dimensional AdS spacetime to the conformal field theory in  $(D-1)$-dimension.  Under this duality, the temperature of the black holes is dual to the temperature of the conformal field theory.
Currently, there are various applications of AdS/CFT duality, such as holographic superconductors \cite{Hartnoll:2008vx,Cai:2015cya}, holographic {Fermi/non-Fermi}  liquids \cite{Lee:2008xf,Liu:2009dm,Cubrovic:2009ye}, and so on \cite{Ryu:2006bv, Balasubramanian1,Cai:2012sk,Cai:2012nm,Johnson:2013dka,Bai:2014tla,Ling:2015dma,Zeng:2015tfj,Gb,Gb2,Zeng:2014xpa}.
Another  possible reason that the AdS space is popular is that the phase structures of the  AdS sapcetime is more abundant, such as  the Hawking-Page phase transition \cite{Hawking83}, Van der Waals-like phase transition \cite{Dolan:2010ha,Cvetic:2010jb}. The Van der Waals-like phase transition exists in the extended phase space, where the negative cosmological constant is treated as the pressure while its conjugate acts as the thermodynamical volume in the Einstein gravity. Imposing the cosmological constant as a dynamical variable, the mass of the black hole corresponds to the enthalpy of the black hole system. The Smarr relation and  the first law of thermodynamics will hold in this case\cite{Kastor:2009wy}.

In this  paper, we will investigate the  laws of thermodynamics and the weak cosmic censorship conjecture in the extended phase space  in the Born-Infeld-anti-de Sitter spacetime.
The first law and phase transition of the Born-Infeld-anti-de Sitter  black hole   have  been investigated extensively. Fernando discussed  its thermodynamics and stability in the grand canonical ensemble \cite{Fernando:2006gh}. Myung investigated its phase transition soon after\cite{Myung:2008eb}. The thermodynamics and phase transition   were  also  investigated from the   point of view  of geometry \cite{Chen:2011hr}. Recently, there have been many works to study the thermodynamics of the Born-Infeld-anti-de Sitter  black hole in the extended phase space, for in this spacetime  there are abundant phases, such as Van der Waals phase transition \cite{Gunasekaran} and reentrant phase transition \cite{Dehyadegari:2017hvd, Zhang:2017lhl}. However, until now, there is still lack of work to discuss the second law  of the thermodynamics  of the Born-Infeld-anti-de Sitter  black hole in the extended phase space. In the Einstein gravity  with a negative cosmological constant, a $D$-dimensional charged AdS black hole with consideration of the pressure and volume
has been investigated \cite{Gwak:2017kkt},  and it was  found that the first law of the thermodynamics holds while the second law is violated for the extremal and near-extremal black holes.  They also discussed the weak cosmic censorship conjecture and found that the extremal and near-extremal black holes do not change their configurations,   so the weak cosmic censorship conjecture is  valid.
In this paper, we will investigate the first law and second law of thermodynamics as well as the  weak cosmic censorship conjecture of  the Born-Infeld-anti-de Sitter black hole in the extended phase space.

Our motivation is two-folded. On one hand,  we   are going to  explore the effect of  the other extensive quantities on the   laws of thermodynamics  and  weak cosmic censorship conjecture besides the
pressure and volume.  In \cite{Gunasekaran}, it was found that  in order to  satisfy the Smarr relation in the Born-Infeld-anti-de Sitter  black hole, the  Born-Infeld parameter should be treated as
a dynamical variable with a conjugate quantity, called  Born-Infeld vacuum polarization. In this case, the first law of thermodynamics is modified   due to  the contribution of the vacuum polarization energy. We want to explore whether the Born-Infeld parameter affects the  laws of thermodynamics  and  weak cosmic censorship conjecture. As a result, the first law and  weak cosmic censorship conjecture are found   not to be  affected.  However,  the second law of the thermodynamics will be affected for the second law is violated for the double horizon black holes and part of the single horizon black holes.
 On the other hand, in  \cite{Gwak:2017kkt}, some approximations are used to investigate the weak cosmic censorship conjecture. The author found that  the extremal and near-extremal black holes did not change their configurations under the   absorption of a charge particle, which is different from that in the normal phase space where the extremal black holes will  deform  into non-extremal black holes \cite{Gwak:2015fsa,Gwak:2016gwj}. In this paper, we intend to find an analytical method without any approximation to study the  configurations of the black holes under a charged particle absorption. We give a general way to check whether the phase space will affect the configurations of black holes as a charged particle is absorbed. Our results show that configurations of the black holes
 will be changed in both the  normal and extended phase space  under a charged particle absorption.   In particular, in both cases the extremal black holes will change into non-extremal black holes.

This paper is organized as follows. In section~\ref{sec2}, the Born-Infeld-anti-de Sitter  black hole is introduced, and the motion of a charged particle around the black hole is investigated. In section~\ref{sec3}, we establish the first law of thermodynamics under the charged particle absorption in the extended phase space, and further discuss the second law of thermodynamics as well as the weak cosmic censorship conjecture. The  second law is violated for the double horizon black holes and part of the single horizon black holes, and the weak cosmic censorship conjecture is   valid  for all black holes.
 In section~\ref{sec4}, we investigate the first law  and second law as well as  weak cosmic censorship conjecture in the normal phase space. We find that for all black holes,  the first law as well as  second law of thermodynamics hold, the weak cosmic censorship conjecture does not violate.
Throughout this paper, we will set the gravitational constant $G$ and the light velocity $c$  to be one.

\section{Motion of a charged particle in the Born-Infeld-anti-de Sitter  black hole}
\label{sec2}

\subsection{ Brief review of the Born-Infeld-anti-de Sitter  black hole}
\label{quintessence_Vaidya_AdS}
The Einstein-Born-Infeld theory in AdS is described by the action \cite{Born1934}
\be
S=\int d^4 x \sqrt{-g}\left[\frac{R-2 \Lambda }{16 \pi G} +\frac{b^2}{4 \pi G}(1-\sqrt{1+\frac{2F}{b^2}})\right],\label{ac}
\ee
in which $F=\frac{1}{4}F_{\mu\nu}F^{\mu\nu}$, $R$ is scalar curvature, $G$ is the gravitational constant,
 $\Lambda$ is the  cosmological constant with $\Lambda=-3/l^2$ where  $l$ is the AdS radius,  and $b$ is the Born-Infeld parameter which relates to the string tension $\alpha$ with the relation  $b = 1/(2 \pi \alpha) $.
The solution of the Born-Infeld AdS black hole can be written as \cite{Fernando2003,Dey2004,Cai2004}
\begin{equation}
 ds^{2}=-f(r)dt^{2}+f^{-1}(r)dr^{2}+r^{2}(d\theta^2+\sin^2\theta d\phi^2),\label{metric1}
\end{equation}
 where%
\begin{equation}
 f(r)=\frac{4 Q^2 \, _2F_1\left(\frac{1}{4},\frac{1}{2};\frac{5}{4};-\frac{Q^2}{b^2 r^4}\right)}{3 r^2}+\frac{ 2 b^2 r^2}{3}
 \left(1-\sqrt{\frac{Q^2}{b^2 r^4}+1}\right)+\frac{r^2}{l^2}-\frac{2 M}{r}+1,\label{metric}
\end{equation}%
in which $M$ represents the ADM mass and $Q$ the electric charge, $_2F_1$
 is the hypergeometric function. From Eq.(\ref{metric}), we know that in the limit $b \rightarrow \infty$, $Q\neq0$,  the  solution reduces to  the Reissner-Nordstr\"{o}m-AdS  black hole, and in the limit $Q\rightarrow 0$, it reduces to the Schwarzschild  AdS  black hole.
The nonvanishing component of the vector potential  is
\begin{equation}
A_t=-\frac{Q_2F_1\left(\frac{1}{4},\frac{1}{2};\frac{5}{4};-\frac{Q^2}{b^2 r^4}\right)}{r},
\end{equation}
which reminds us  that the chemical potential depends on the Born-Infeld parameter.

In \cite{Myung:2008eb, Banerjee:2011cz}, the author   computed  the horizon of the   extremal  black hole   as
\begin{equation}
r_{e}^2 = \frac{l^2}{6}\left(\frac{1 + \frac{3}{2b^2 l^2}}{1 +
\frac{4}{4b^2 l^2}}\right) \left[- 1 + \sqrt{ 1 + \frac{12\left(1
+ \frac{3}{4b^2 l^2}\right)}{b^2l^2\left(1 + \frac{3}{2b^2 l^2}
\right)^2} \left(b^2Q^2 -\frac{1}{4} \right)}~\right].
\end{equation}
They claimed that  $ b Q \ge 0.5$ should be satisfied in order to have a
real root for $r^2_{e}$.   In particular,  they stressed that  $ 0 \le b Q < 0.5$ is  a forbidden region for in this region there is  no black hole solution. However, it was later
found that there   was  a single horizon black hole solution  in the region  $ 0 \le b Q < 0.5$ \cite{Cai2017prd}. The action growth of the Wheeler-DeWitt  patch  for the single horizon black hole has been calculated and it was found  that the Lloyd bound   was  satisfied. In fact, the single horizon black hole exists not only in the region
$ 0 \le b Q < 0.5$, but also in  $ b Q \ge 0.5$, please   refer to  Figure (\ref{fig3}) or  Figure (\ref{fig5}).   Black holes  having
  two horizons exist only in a narrow region of the value of $M$.
  In \cite{Gunasekaran}, the author obtained a marginal mass
  \be \label{mm}
  M_{m}=\frac{1}{6}\sqrt{\frac{b}{\pi}}Q^{3/2}\Gamma(\frac{1}{4})^2.
  \ee
   If the black hole mass  is larger than the marginal mass, there are only single horizon black holes.

\subsection{  Energy and momentum of a particle absorbed by a black hole }

In this subsection, we will consider the   dynamics  of a charged particle as it is absorbed by the Born-Infeld-anti-de Sitter black hole. We   will  concentrate mainly on the  relations between the conserved quantities,  such as the energy and the momentum.

 The Hamilton-Jacobi equation for the vector potential $A_\mu$ is
\be \label{HJE}
g^{\mu\nu}(P_\mu-e A_\mu)(P_\nu-e A_\nu)+u^2=0,
\ee
in which $u$ is the rest mass, $e$ is the  electric  charge, and
$P_\mu$ is the momentum of the particle   with the definition
\be \label{eq:pmu}
P_\mu = \partial_\mu {\cal I},
\ee
where  ${\cal I}$ is the Hamilton-Jacobi action. Considering the  symmetry  of the black hole, the action can be written as
\be \label{eqI}
{\cal I} = - E t  +{\cal I}_r(r) + {\cal I}_\theta(\theta) +L \phi,
\ee
where $E$ and $L$ are the energy and angular momentum of the particle, respectively.  They are the conserved quantities  with respect to $t$ and $\phi$.  From  Eq.\eqref{metric1}, the inverse of the metric is,
\be
g^{\mu\nu}\partial_\mu\partial_\nu = -f(r)^{-1}(\partial_t)^2+f(r)(\partial_r)^2
+ r^{-2} (\partial_\theta ^2 + \sin^{-2}\theta \partial_\phi^2).\label{eq126}
\ee
 Therefore, the Hamilton-Jacobi  equation can be re-expressed as
\be
u^2-f(r)^{-1}(-E-eA_t)^2+f(r)(\partial_r {\cal I}_r(r))^2
+ r^{-2} ((\partial_\theta {\cal I}_\theta)^2 + \sin^{-2}\theta L^2)=0. \label{eqHJeq}
\ee
 One can readily separate the angular part from the above Eq.\eqref{eqHJeq} and define it as
\be
\quad K =  (\partial_\theta {\cal I}_\theta)^2 +\frac{1}{ \sin^2\theta}L^2,
\ee
in which $K$ can be solved from the radial part in Eq.\eqref{eqHJeq} as
\be
K = - u^2 r^2 +\frac{ r^2}{f(r)}(-E-eA_t)^2
-r^2f(r)(\partial_r {\cal I}_r(r))^2.
\ee
Therefore,  Eq.(\ref{eqI}) can be rewritten as
\be \label{eq:I02}
\mathcal{I}=-Et +\int dr \sqrt{R}  + \int d\theta \sqrt{\Theta}  +L\phi,
\ee
 with
\bea
\mathcal{I}_r &\equiv& \int dr \sqrt{R},\quad \mathcal{I}_\theta \equiv \int d\theta \sqrt{\Theta},\quad\Theta=K-\frac{1}{\sin^2\theta}L^2, \nonumber\\
R&=&\frac{1}{ r^2 f(r)}\left(-K- u^2 r^2\right)+\frac{1}{ r^2 f(r)}\left(\frac{ r^2}{f(r)}(-E-eA_t)^2\right).
\ena
 Hence, from  Eq.(\ref{eq:I02}),  the radial momentum  $P^r$ and angular momentum $P^\theta$ can be written as
\be
P^r  =f(r) \sqrt{ -\frac{K+u^2 r^2}{ r^2 f(r)} +\frac{1}{f^2(r)}(-E-eA_t)^2}, \label{eq:pr}
\ee
\be
P^\theta = \frac{1}{r^2 } \sqrt{K -\frac{1}{ \sin^2\theta}L^2}.
\ee
We will   study  how the black hole  thermodynamics changes as  a  charged particle is absorbed by the black hole. Concretely, we focus on the relation between the momentum and energy. In principle, as $K$ is eliminated, we can get  their relation at any locations. Our goal is to investigate the thermodynamics on the event horizon, thus we mainly pay attention to the near horizon behavior of the particle. In this case, Eq.(\ref{eq:pr}) can be simplified as
\be \label{eq:dispersion01}
E=\frac{Q_2F_1\left(\frac{1}{4},\frac{1}{2};\frac{5}{4};-\frac{Q^2}{b^2 r_+^4}\right)}{r_+}e+|P^r_+|.
\ee
 It should be stressed that a positive sign should be endowed in front of the $|P^r_+|$ term.  This choice is to assure that the signs in front of $E$ and $|P^r_+|$ are the same and positive in the positive flow of time \cite{Christodoulouprl}. From Eq.(\ref{eq:dispersion01}), we know that the energy dependents on the electric potential too. However, the potential is independent of the flow of time and only related to the interaction between particle and black hole. Thus, the total value of energy under the sum of the potential is not important, and we simply choose a positive sign in front of $|P^r_+|$.

\section{Thermodynamics and weak  cosmic  censorship  conjecture in the normal phase space }
\label{sec4}
In this section, we   will examine   whether the first law,  the second law  as well as the weak cosmic cosponsorship conjecture are valid  in the normal phase space of Born-Infeld-anti-de Sitter spacetime. Our   discussions are mainly  based on the relation between the energy and momentum of the absorbed particle in Eq.(\ref{eq:dispersion01}).
 %Ourmotivation is to see the differences of the  thermodynamic laws and weak cosmic cosponsorship conjecture in different phase spaces.

\subsection{Thermodynamics in the normal phase space}

In the Born-Infeld-anti-de Sitter  black hole, the electrostatic potential difference between the black hole horizon and the infinity is
\begin{equation} \label{phi}
\Phi=\frac{Q_2F_1\left(\frac{1}{4},\frac{1}{2};\frac{5}{4};-\frac{Q^2}{b^2 r_+^4}\right)}{r_+},
\end{equation}
in which $r_+$ is the event horizon of the black hole, which is determined from $f(r_+)=0$.
The Hawking temperature,  defined by $T=\frac{f^{\prime}(r)}{4\pi}\mid_{r_+}$, can be written as
\begin{equation}
T=\frac{-2 l^2 Q^2 \, _2F_1\left(\frac{1}{4},\frac{1}{2};\frac{5}{4};-\frac{Q^2}{b^2 r_+^4}\right)+r_+^4 \left(3-2 b^2 l^2 \left(\sqrt{\frac{Q^2}{b^2 r_+^4}+1}-1\right)\right)+3 l^2 M r_+}{6 \pi  l^2 r_+^3},\label{eq13}
 \end{equation}
In addition, according to the  Bekenstein-Hawking  entropy area relation, we can get the black hole entropy
\begin{equation}
S=\pi r_+^2.  \label{eq15}
 \end{equation}
At the horizon, the mass $M$ can be expressed as
\be \label{mass23}
M=\frac{4 l^2 Q^2_2F_1 \left[\frac{1}{4},\frac{1}{2},\frac{5}{4},-\frac{Q^2}{b^2 r_+^4}\right]+3 l^2 r_+^2+3 r_+^4+2 b^2 l^2 r_+^4-2 b^2 l^2 \sqrt{1+\frac{Q^2}{b^2 r_+^4}} r_+^4}{6 l^2 r_+}.
\ee
With Eqs.(\ref{phi}), (\ref{eq13}), (\ref{eq15}), (\ref{mass23}),  we can obtain    the first law  of thermodynamics.

In the normal phase space, the cosmological parameter is a constant, and the mass $M$ is the internal energy of the black hole. According to the energy conservation and charge conservation, as a
 charged particle is absorbed by the black hole, the variation of the internal energy and charge  satisfy
\be
E=dM,  e=dQ,
\ee
so that Eq.(\ref{eq:dispersion01}) should be written as
\be \label{eq:dispersion023}
dM=\frac{Q_2F_1\left(\frac{1}{4},\frac{1}{2};\frac{5}{4};-\frac{Q^2}{b^2 r_+^4}\right)}{r_+}dQ+|P^r_+|.
\ee
In addition, as the charged particle is absorbed by the black hole, the variation of the  entropy can be written as
\be  \label{eq:s02}
dS=2 \pi  r_{+} dr_{+},
\ee
where we have used Eq.(\ref{eq15}). To obtain the last  result of the variation of the entropy, we should  find $dr_+$ firstly.

The absorbed particle leads to a variation of the event horizon of the black hole, which further leads to  the change of $f(r)$.  The variation of  $f(r)$, labeled by
  $df_\text{+}$, satisfies
\be \label{eq:function053}
df_\text{+}=\frac{\partial f_\text{+}}{\partial M}dM+\frac{\partial f_\text{+}}{\partial Q}dQ+\frac{\partial f_\text{+}}{\partial r_\text{+}}dr_\text{+}=0.
\ee
  Substituting Eq.(\ref{eq:dispersion023}) into Eq.(\ref{eq:function053}), we can get  $dr_{+}$  directly as
 \bea\label{eq:variables0233}
dr_{+}=\frac{-3 l^2 |P^r_+| r_+^2}{2 l^2 Q^2_2F_1\left[\frac{1}{4},\frac{1}{2},\frac{5}{4},-\frac{Q^2}{b^2 r_+^4}\right]-3r_+^4-3 M l^2 r_++2 b^2 l^2  \left(-1+\sqrt{1+\frac{Q^2}{b^2 r_+^4}}\right) r_+^4}.
\ena
 With Eq.(\ref{eq:s02}), the variation of entropy can be expressed as
\bea\label{eq:s0233}
dS=\frac{-6\pi l^2 |P^r_+| r_+^3}{2 l^2 Q^2_2F_1\left[\frac{1}{4},\frac{1}{2},\frac{5}{4},-\frac{Q^2}{b^2 r_+^4}\right]-3r_+^4-3 M l^2 r_++2 b^2 l^2  \left(-1+\sqrt{1+\frac{Q^2}{b^2 r_+^4}}\right) r_+^4}.
\ena
  Combining  Eqs.(\ref{eq13}) and (\ref{eq:s0233}), we find $T dS=|P^r_+|$. Therefore the internal energy in Eq.(\ref{eq:dispersion023}) can be rewritten as
\begin{align}
dM=TdS+\Phi dQ,
\end{align}
which is  the first law of the black hole thermodynamics in the normal phase space.

Next we turn to study the second law of the thermodynamics, which states that the entropy of the black hole never decreases in the  arrow of time. As the charged particle is absorbed by the black hole, the entropy of the black hole increases according to the second law of the black hole thermodynamics. We will employ Eq.(\ref{eq:s0233}) to check whether this is true in the normal phase space.

We first discuss the extremal  black hole, for which the inner horizon and outer horizon coincide and the temperature vanishes at the horizon.  With Eq.(\ref{eq13}), we can get the mass of the extremal black hole
\be \label{mass}
M_e=\frac{2 l^2 Q^2 \, _2F_1\left(\frac{1}{4},\frac{1}{2};\frac{5}{4};-\frac{Q^2}{b^2 r_+^4}\right)+2 b^2 l^2 r_+^4 \sqrt{\frac{Q^2}{b^2 r_+^4}+1}-2 b^2 l^2 r_+^4-3 r_+^4}{3 l^2 r_+}.
\ee
Substituting  Eq.(\ref{mass}) into Eq.(\ref{eq:s0233}), we get
\be\label{eq:variables00013}
dS_{extreme}=\infty.
\ee
The divergency of the variation of the entropy  means that it is meaningless to investigate the second law of thermodynamics for the extremal black holes. Therefore we concentrate on mainly the non-extremal black holes thereafter.

\begin{center}
{\footnotesize{\bf Table 1.} Numerical results of the variation of entropy in the normal phase space for the case of $Q=1, b=0.8$.\\
\vspace{2mm}
\begin{tabular}{ccc}
\hline
{M}         &{$r_{+}$}      & {dS}            \\    \hline
{1.0302893} &  {0.377831}  &{7951.91}         \\
{1.06}     & {0.559903}   &{10.1647}      \\
{1.09}     & {0.630175}  &{7.75712}     \\
{1.12}      &  {0.682279} &{6.67212}      \\
{1.15}    &  {0.725166}  &{6.01545}            \\
{1.18}   & {0.762249}     &{5.56131}            \\
{1.21}   & {0.795261}   &{5.22208}          \\
{1.23}    &{0.815527}  &{4.95557}          \\
{1.26}    &  {0.843831}  &{4.80642}            \\
{1.29}   & {0.870061}     &{4.61436}            \\
{1.32}   & {0.894579}   &{4.45151}          \\
{1.35}    &{0.917657}  &{4.31095}          \\
\hline
\end{tabular}}
\end{center}

As the charge $Q$ and the Born-Infeld parameter $b$  are given, we can find the extremal black hole  mass $M_{e}$ via asymptotic evaluation method\footnote{We plot the relation between $f(r)$ and $r$ for different values of $M$ as the other parameters are given. For the case that there is only a solution for $f(r)$, the corresponded mass is the extremal mass.  }. With the extremal mass, we also can get the extremal radius via Eq.(\ref{mass}). Then we can
give any mass, which should be larger than the critical mass, to find the  corresponding  horizon and variation of the entropy of the non-extremal black hole. Throughout this paper, we set $l=P^r_+=1$. For the case $Q=1, b=0.8$, we find that the extremal black hole  mass is $M_{e}=1.0302893$.
For the extremal and non-extremal black holes, the variations of entropy and horizon $r_h$ are listed in Table 1. We can see that for  all the black holes, the variations of entropy are positive. That is to say, the second law of black hole thermodynamics always holds as a charged particle is absorbed by  the black hole in the normal phase space.

\begin{figure}[H]
\centering
\includegraphics[scale=0.65]{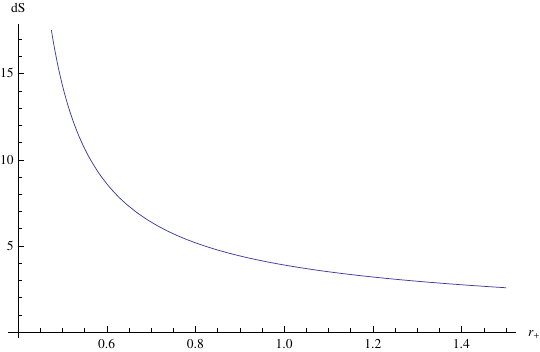}
 \caption{\small The relation between  $dS$  and $r_+$ for the case of $Q = 1, b = 0.8$ in the normal phase space  } \label{fig4}
\end{figure}

Substituting Eq.(\ref{mass23}) into Eq.(\ref{eq:s0233}), the variation of the entropy can be rewritten as
\be \label{ds3}
dS=-\frac{4 l^2 \pi  r_+}{-3 r_+^2+l^2 \left(-1+2 b^2 \left(-1+\sqrt{1+\frac{Q^2}{b^2 r_+^4}}\right) r_+^2\right)}.
\ee
On the basis of  Eq.(\ref{ds3}), we can plot the relation between  the variation of the entropy and the horizon, which is shown in Figure (\ref{fig4}). One can see that the variation of entropy
is positive for all black holes, which is consistent with the results in Table 1.

\begin{center}
{\footnotesize{\bf Table 2.} Numerical results of the variation of entropy in the normal phase space for the case of $Q=0.8, b=0.8$.\\
\vspace{2mm}
\begin{tabular}{ccc}
\hline
{M}         &{$r_{+}$}      & {dS}            \\    \hline
{0.7671287} &  {0.257154}  &{13834.3}         \\
{0.77}   & {0.327056}   &{24.1034}          \\
{0.78}    &{0.399628}  &{12.9661}          \\
{0.79}     & {0.443068}   &{10.392}      \\
{0.80}   & {0.476571}     &{9.08111}            \\
{0.81}     & {0.504592}  &{8.24664}     \\
{0.82}      &  {0.529029} &{7.6531}      \\
{0.83}    &  {0.550894}  &{7.20171}            \\
{0.84}   & {0.570803}     &{6.84265}            \\
{0.85}   & {0.589163}   &{6.5477}          \\
\hline
\end{tabular}}
\end{center}

\begin{figure}[H]
\centering
 \includegraphics[scale=0.65]{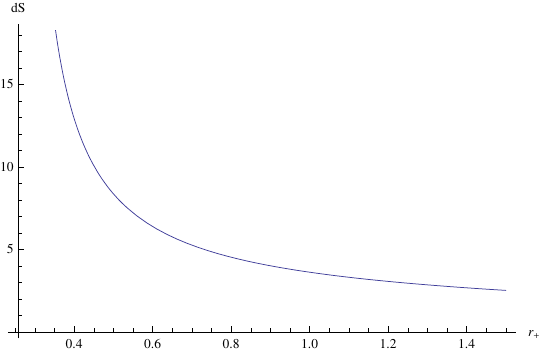}
 \caption{\small The relation between  $dS$  and $r$ for the case of $Q = 0.8, b = 0.8$ in the normal phase space.   }
 \label{fig6}
\end{figure}

To confirm that our conclusion is  independent of  $Q$ and $b$, we can choose some other values of them.
 For the case of $Q=0.8, b=0.8$, the extremal mass of the black hole is 0.7671287. For the extremal and non-extremal black holes, the variations of entropy are listed in Table 2,  and the  relation between  the variations of  entropy and the horizons are plotted in Figure (\ref{fig6}).
 From them, we can see that the variation of entropy is positive too. In other words, the second law of black hole thermodynamics is satisfied in the normal phase space for the
 Born-Infeld-anti-de Sitter  black hole, which is independent of the  charge and Born-Infeld parameter.

 In addition, form Eq.(\ref{mm}), we know that as the mass of the black hole is larger than the critical mass $M_m$, the black hole are single horizon black holes. For the cases $Q=1, b=0.8$ and $Q=0.8, b=0.8$, the critical mass are $M_m=1.10556$  and $M_m=0.791072$ respectively.  Table 1 and Table 2 show that the second law   is satisfied in the normal phase space not only for the double horizon black holes  but also for the single horizon black holes.
\subsection{Weak cosmic censorship conjecture in the normal phase space}

The weak cosmic censorship conjecture   asserts that there is no singularities  visible from future null infinity. In other words, singularities need to be hidden from an observer at infinity by the event horizon of a black hole. So an event horizon should exist to assure   the validity of the  weak cosmic censorship conjecture.  As a charged particle is absorbed by the black hole, we will check whether there is an event horizon. For the single horizon black holes in the Born-Infeld theory, the horizons will not be  broken while for the double horizon black holes, they may be, which can be seen from Figure (\ref{fig3}). The  weak cosmic censorship conjecture thus is valid for the single horizon black holes always and we only should discuss the double horizon black holes thereafter.

 For a double event horizon black hole, there is a minimum value  for $f(r)$, the corresponding radial coordinate is labeled as  $r_{m}$. At $r_{m}$, we find there are always
 the following relations
\bea \label{condition}
&&f|_{r=r_{m}}\equiv f_{m}=\delta\leq 0, \nonumber\\
&&\partial_{r}f|_{r=r_{m}}\equiv f'_{m}=0,\nonumber \\
&&(\partial_{r})^2 f|_{r=r_{m}}\equiv f^{\prime\prime}_{m}>0.
\ena
For the extremal black hole, $\delta=0$, and for the near extremal black hole, $\delta$ is a small quantity. For the extremal and near-extremal black holes, as a charged particle is absorbed by the black hole, $f_{m}$ may move upward or downward, which correspond to that there  does not exist or exist horizon  respectively. Next, we will find how $f_{m}$   move in the normal phase space.

 The absorption of a charged particle will lead to  variations of the
 mass  and  charge  of the black hole. Correspondingly, the locations of
the minimum value and event horizon will change as $r_{m}\rightarrow r_{m}+dr_{m}$, $r_{+}\rightarrow r_{+}+dr_{+}$.
The variation of $f(r)$, defined by $df_{m}$,  can be expressed  as
\be  \label{eqc13}
df_{m}=\left(\frac{\partial f_{m}}{\partial M}dM+\frac{\partial f_{m}}{\partial Q}dQ\right),
\ee
where we have used $f'_{m}=0$. In addition, at the new  minimum point, there is also a relation
\begin{align} \label{eqc0}
\partial_{r} f|_{r=r_m+dr_m}
=f'_{m}+df'_{m}=0,
\end{align}
 which implies
\be\label{eqc23}
df'_{m}=\frac{\partial f'_{m}}{\partial M}dM+\frac{\partial f'_{m}}{\partial Q}dQ+\frac{\partial f'_{m}}{\partial r_{m}}dr_{m}=0.
\ee
With  the condition
$f'_{m}=0$, we can obtain  $M$ and further $dM$, which is
\bea \label{eqc33}
&dM&=r_m^2 \left(\frac{2 b^4 dr \sqrt{1+\frac{Q^2}{b^2 r_m^4}} r_m^4}{Q^2+b^2 r_m^4}+b^2 \left(-2 dr+\frac{dQ Q \sqrt{1+\frac{Q^2}{b^2 r_m^4}} r_m}{Q^2+b^2 r_m^4}\right)\right)-r_m^2 \frac{3 dr }{l^2} \nonumber\\
&~~~~~+&\frac{dQ Q _2F_1\left[\frac{1}{4},\frac{1}{2},\frac{5}{4},-\frac{Q^2}{b^2 r_m^4}\right]}{r_m}.
\ena
We are   interested in the extremal black holes, for which  Eq.(\ref{eq:dispersion023}) is valid at $r_m$.  Inserting Eq.(\ref{eqc33}) into Eq.(\ref{eq:dispersion023}), we get
\be\label{eqc43}
dr_{m}=\frac{l^2 \left(|P^r_+| \left(b^2 r_m^4+Q^2\right)-b^2 dQ Q r_m^3 \sqrt{\frac{Q^2}{b^2 r_m^4}+1}\right)}{r_m^2 \left(b^2 r_m^4 \left(2 b^2 l^2 \left(\sqrt{\frac{Q^2}{b^2 r_m^4}+1}-1\right)-3\right)+Q^2 \left(-2 b^2 l^2-3\right)\right)}.
\ee
In addition, substituting Eq.(\ref{eqc33}) into  Eq.(\ref{eqc23}),  $dr_{m}=0$ will be produced,  which means
 \be \label{eqc53}
 dQ=\frac{|P^r_+| r_m \sqrt{\frac{Q^2}{b^2 r_m^4}+1}}{Q}.
 \ee
 Similarly,
substituting Eq.(\ref{eqc33}) into Eq.(\ref{eqc13}), $df_{m}$ can be expressed as a function with respect to $dQ, dr_{m}$. From  Eq.(\ref{eqc53}), Eq.(\ref{eqc13}) can be simplified as
\be \label{eqc63}
df_{m}=-\frac{2 |P^r_+|}{r_{m}},
\ee
  which shows that there is a shift of $f_m$ in the negative direction as a charged particle is absorbed by the black hole. In other words, the weak cosmic censorship conjecture holds  in the normal  phase space since there always exist horizons to assure
the singularity of the spacetime to be hidden.

\section{Thermodynamics and weak cosmic censorship conjecture in the extended phase space }
\label{sec3}

 In this section, we will examine  the thermodynamics and   weak cosmic censorship conjecture  in the extended phase space of  Born-Infeld-anti-de Sitter spacetime  under a charged particle absorption with  Eq.(\ref{eq:dispersion01}).

\subsection{Thermodynamics in the extended phase space}

In the extended phase space, the cosmological   constant  is treated as the pressure  $Y$ and the corresponding  conjugate quantity  is treated as the volume $V$. In this case,   in order to satisfy the Smarr relation, the Born-Infeld parameter  $b$  should  also  be treated as  an extensive quantity. The Smarr relation  for the Born-Infeld-anti-de Sitter  black hole is \cite{Gunasekaran}
\be
M=2(T S-V Y)+\Phi Q-Bb, \label{eq19}
\ee
in which
\bea
&Y&=-\frac{\Lambda}{8 \pi}=\frac{3}{8 \pi l^2},\nonumber \\
&V&=\frac{4}{3} \pi r_+^3,\nonumber \\
&B&=\frac{Q^2 _2F_1\left[\frac{1}{4},\frac{1}{2},\frac{5}{4},-\frac{Q^2}{b^2 r_+^4}\right]-2 b^2 \left(-1+\sqrt{1+\frac{Q^2}{b^2 r_+^4}}\right) r_+^4}{3 b r_+}, \label{eq16}
\ena
where $B$, which is the conjugate quantity of $b$, is regarded as the Born-Infeld vacuum polarization \cite{Gunasekaran}.
It should be stressed that $M$ is not the internal energy but the  enthalpy of the thermodynamic system,  which relates to the   internal energy with the following relation \cite{Gunasekaran}
\be
M=U+YV+bB. \label{eq119}
\ee
 As the charged particle is absorbed by the black hole, the energy and charge are supposed to be conserved. Namely the energy and charge of the particle equal to the varied energy and charge of the black hole.  According to Eq.(\ref{eq:dispersion01}), we know that the energy of the particle is determined by the charge and radial momentum of the particle near the event horizon. Our goal is to obtain the first law of thermodynamics, we thus should find some quantities which can be expressed by the charge  and radial momentum  of the particle.

 Based on the energy conservation and charge conservation, we can obtain
\be
E=dU=d(M-YV-bB),  e=dQ,
\ee
 The energy in Eq.(\ref{eq:dispersion01}) changes  accordingly  into
\be \label{eq:dispersion02}
dU=\frac{Q_2F_1\left(\frac{1}{4},\frac{1}{2};\frac{5}{4};-\frac{Q^2}{b^2 r_+^4}\right)}{r_+}dQ+|P^r_+|.
\ee
The variation of the event horizon of the black hole, denoted by $dr_{+}$, is determined by the charge, energy, and radial momentum of the absorbed particle, leading to the changes of $f(r)$.  However,   near event horizon,   $df(r_{+})\equiv df_\text{+}$ will  not change since  $f(r_++dr_{+})=0$, that is
\be \label{eq:function05}
df_\text{+}=\frac{\partial f_\text{+}}{\partial M}dM+\frac{\partial f_\text{+}}{\partial Q}dQ+\frac{\partial f_\text{+}}{\partial l}dl+\frac{\partial f_\text{+}}{\partial r_\text{+}}dr_\text{+}+\frac{\partial f_\text{+}}{\partial b}db=0.
\ee
In addition, with the help of  Eq.(\ref{eq119}), Eq.(\ref{eq:dispersion02}) can be expressed as
\be \label{eq:dispersion03}
dM-d(YV+Bb)=\frac{Q_2F_1\left(\frac{1}{4},\frac{1}{2};\frac{5}{4};-\frac{Q^2}{b^2 r_+^4}\right)}{r_+}dQ+|P^r_+|.
\ee
  From Eq.(\ref{eq:function05}), we can obtain $dl$, and substituting  $dl$ into Eq.(\ref{eq:dispersion03}), we can delete it directly. Interestingly, $dQ$, $db$, and $dM$ are also eliminated at the same time. In this case, there is only a  relation between $|P^r_+|$  and  $dr_{+}$,  which is
 \be \label{eq:variables02}
dr_\text{+}=\frac{-6 l^2 r_+^2(|P^r_+|+b dB)}{4 l^2 Q^2 \, _2F_1\left(\frac{1}{4},\frac{1}{2};\frac{5}{4};-\frac{Q^2}{b^2 r_+^4}\right)+l^2 \left(4 b^2 r_+^4 \left(\sqrt{\frac{Q^2}{b^2 r_+^4}+1}-1\right)-6 M r_+\right)+3 r_+^4}.
\ee
  Therefore,  the variations of entropy and volume of the black hole can be expressed as
\be\label{eq:variables001}
dS=\frac{-12 \pi  l^2 (|P^r_+|+b dB) r_+^3}{4 l^2 Q^2 \, _2F_1\left(\frac{1}{4},\frac{1}{2};\frac{5}{4};-\frac{Q^2}{b^2 r_+^4}\right)+r_+^4 \left(4 b^2 l^2 \left(\sqrt{\frac{Q^2}{b^2 r_+^4}+1}-1\right)+3\right)-6 l^2 M r_+},
\ee
\be\label{eq:variables002}
dV=\frac{-24 \pi  l^2 (|P^r_+|+b dB) r_+^4}{4 l^2 Q^2 \, _2F_1\left(\frac{1}{4},\frac{1}{2};\frac{5}{4};-\frac{Q^2}{b^2 r_+^4}\right)+r_+^4 \left(4 b^2 l^2 \left(\sqrt{\frac{Q^2}{b^2 r_+^4}+1}-1\right)+3\right)-6 l^2 M r_+}.
\ee
Based on   the above formulae, we   reach
\be
T dS-YdV-b dB=|P^r_+|.
\ee
The internal energy in Eq.(\ref{eq:dispersion02}) thus will change into
\be \label{eq:variables003}
dU=\Phi dQ + T dS-YdV-b dB.
\ee
In the extended phase space, the mass of the black hole have been defined as   the  enthalpy. From Eq.(\ref{eq119}), we can get the relation between the enthalpy and internal energy  as
\be \label{eq:variables004}
dM=d U+Y dV+VdY+Bdb+bdB.
\ee
Substituting Eq.(\ref{eq:variables004}) into Eq.(\ref{eq:variables003}), we get
\be
dM=TdS+\Phi dQ+VdY+Bdb,
\ee
which is obviously the first law of the black hole thermodynamics in the extended phase space \cite{Gunasekaran}. Namely,  the first law of thermodynamics holds in the extended phase space of Born-Infeld-anti-de Sitter spacetime under a charged particle absorbtion.

Next we will focus on the second law of the thermodynamics in the extended phase space with Eq.(\ref{eq:variables001}). It should be stressed that in Eq.(\ref{eq16}), the variation $dB$  is the function of  $dr_+, dQ, db$. However, the existence of  $dQ, db$ would affect the definition of $\Phi$ and  $B$ respectively. Thus, $dQ, db$ may relate to  $dr_+$, resulting in that $dB$ is only a function of  $dr_+$. Without loss of generality,  we  will regard $dB$ as $g h dr_+$, in which $g\equiv \partial B/ \partial r_+$ and $h$ are the contributions of $\partial B/ \partial Q$ and $\partial B/ \partial b$.  For simplicity, we 
will set $k=gh$
thereafter, where $k$ is a positive parameter for the energy of the black hole increases as a charged particle is absorbed by the black hole\footnote{Here we can not determine the value of $k$ by the parameters of black holes for we do not know the value $h$ though we know $g$.}. In this case,  Eq.(\ref{eq:variables001})  would change into
\be
dS=\frac{-12 l^2 \pi  |P^r_+| r_+^3}{3 r_+^4-6 M l^2r_+ +2 b l^2r_+^2 \left(3 k+2 b \left(-1+\sqrt{1+\frac{Q^2}{b^2 r_+^4}}\right) r^2\right)+4 l^2 Q^2 _2F_1\left[\frac{1}{4},\frac{1}{2},\frac{5}{4},-\frac{Q^2}{b^2 r_+^4}\right]}. \label{eqev1}
\ee
We will use Eq.(\ref{eqev1}) to check the second law of thermodynamics.

For the extremal black hole,  as Eq.(\ref{mass}) is substituted into  Eq.(\ref{eqev1}), we obtain
\be\label{eq:variables0001}
dS_{extreme}=-\frac{4 \pi  l^2 |P^r_+|  r_e}{2 b k l^2+3 r_e^2}<0,
\ee
in which $r_e$ is the radius of horizon  of the extremal black hole. It is obvious that the variation of the entropy is  always negative. In other words, the entropy of the black hole decreases and the second law of black hole thermodynamics is  violated  for the extremal Born-Infeld-anti-de Sitter  black hole.

Now we turn to the non-extremal black hole.  In Table 3, the variation of entropy for different $k$'s are given.  For the case of $k=0.01$, we find that the variation of the entropy is negative for $M\leq 1.18$ while it is positive for $M>1.18$. As $k$ increases to 0.05, the dividing point, which
  determines  the negative or positive of $dS$, jumps to 1.29.

\begin{center}
{\footnotesize{\bf Table 3.} Numerical result of the variation of entropy in the extended phase space for different $k$'s with $Q=1, b=0.8$.\\
\vspace{2mm}
\begin{tabular}{ccccc}
\hline
{M}         &{$r_{+}$}      & {dS(k=0.01)}      & {dS(k=0.05)}     & {dS(k=0.1)}   \\    \hline
{1.0302893} &  {0.377831}  &{-10.7015}       &{-9.35242}     &{-8.07927}                                       \\
{1.06}     & {0.559903}   &{-26.6229}      &{-21.4326}  &{-17.2331}                                                  \\
{1.09}     & {0.630175}  &{-44.7806}    &{-31.614}     &{-23.3659}                                      \\
{1.12}      &  {0.682279} &{-67.2451 }    &{-44.7716}     &{-31.5792}     \\
{1.15}    &  {0.725166}  &{-115.771}        &{-63.8534}    &{-40.9168}    \\
{1.18}   & {0.762249}     &{-261.085}      &{-95.1325}    &{-53.0124}       \\
{1.21}   & {0.795261}   &{25611.5}         &{-157.107}   &{-69.5882}    \\
{1.23}    &{0.815527}  &{446.991}       &{-249.513}       &{-84.645}  \\
{1.26}    &  {0.843831}  &{196.217}        &{-1064.83}     &{-117.876}    \\
{1.29}   & {0.870061}     &{132.636}       &{593.162}      &{-177.586}    \\
{1.32}   & {0.894579}   &{103.576}          &{252.42}   &{-316.98}  \\
{1.35}    &{0.917657}  &{86.9155}          &{167.913}  &{-1018.35}     \\
\hline
\end{tabular}}
\end{center}

\begin{figure}[H]
\centering
\includegraphics[scale=0.65]{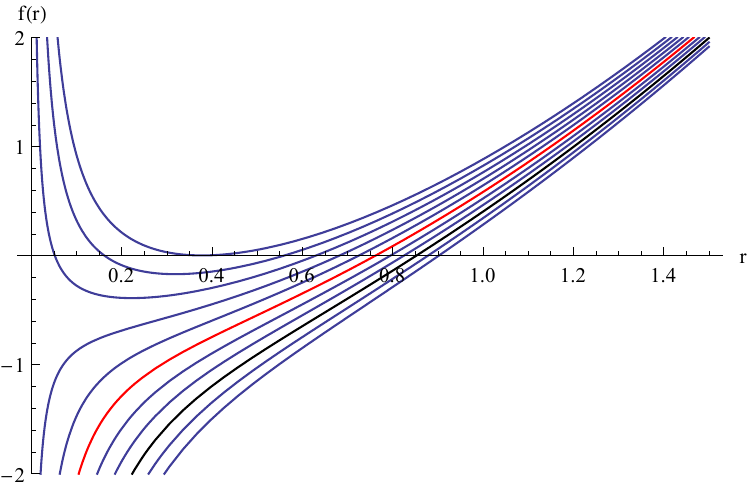}
 \caption{\small The relation between  $f(r)$  and $r$ for the case $Q = 1, b = 0.8$. Curves from top to bottom correspond to $M$ varying from 1.03 to 1.35 with
 step 0.03.   } \label{fig3}
\end{figure}

In Figure (\ref{fig3}), we plot the relation between $f(r)$ and $r$. We can see that there exist single horizon black holes and double horizon black holes for different masses $M$. We are interested in the case  $M= 1.18, 1.29$, which has been marked by red and black in Figure (\ref{fig3}). Obviously, for both cases, the black holes are single horizon black holes.  So from Table 3, we can conclude that the variations of entropy for all the  double horizon black holes and part of the single horizon  black holes are negative. In other words, all the double horizon black holes  and part of the single horizon  black holes violate the second law of black hole thermodynamics as a charged particle is absorbed by  the black hole.  As $k$ increases, the mass of the black holes that violate the second law of black hole thermodynamics will be  larger.

\begin{center}
{\footnotesize{\bf Table 4.}   Numerical results  of the variation of entropy in the normal extended space for different $k$'s with $Q=0.8, b=0.8$.\\
\vspace{2mm}
\begin{tabular}{ccccc}
\hline
{M}         &{$r_{+}$}      & {dS(k=0.01)}     & {dS(k=0.05)}     & {dS(k=0.1)}    \\    \hline
{0.7671287} &  {0.257154}  &{-15.0897}    &{-11.6177}        &{-9.0227}   \\
{0.77}   & {0.327056}   &{-24.7011}      &{-17.8392}      &{-13.2413}  \\
{0.78}    &{0.399628}  &{-46.5846}      &{-29.2307}       &{-19.9438} \\
{0.79}     & {0.443068}   &{-80.512}     &{-41.8143}     &{-26.1208}  \\
{0.80}   & {0.476571}     &{-158.078}          &{-58.7798}   &{-32.9262}   \\
{0.81}     & {0.504592}  &{579.968}    &{-84.6206}        &{-40.9266}  \\
{0.82}      &  {0.529029} &{533.602}      &{-129.248}       &{-50.6305} \\
{0.83}    &  {0.550894}  &{198.864}         &{-237.174}     &{-63.4017}    \\
{0.84}   & {0.570803}     &{130.85}           &{-781.178}    &{-80.43}  \\
{0.85}   & {0.589163}   &{100.887}         &{788.852}       &{-104.846}    \\
\hline
\end{tabular}}
\end{center}

\begin{figure}[H]
\centering
 \includegraphics[scale=0.65]{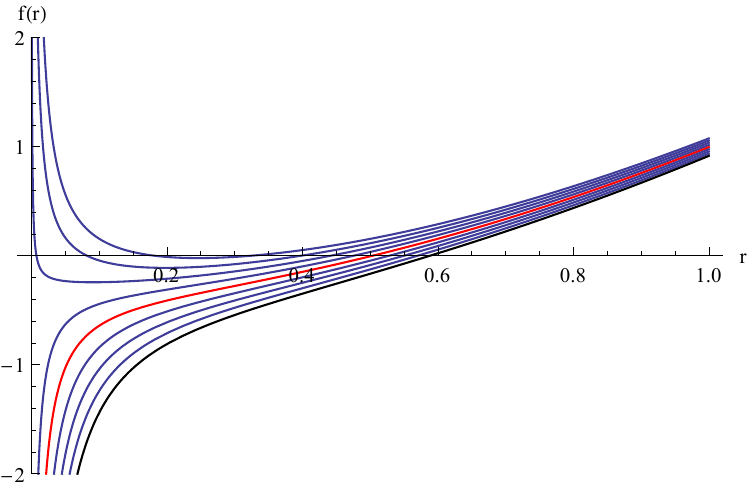}
 \caption{\small The relation between  $f(r)$  and $r$ for the case $Q = 0.8, b = 0.8$. Curves from top to bottom correspond to $M$ varying from  0.77 to 0.85 with
 step 0.01.   }
 \label{fig5}
\end{figure}

  In order to confirm our conclusion, we choose different  charges and  Born-Infeld parameters  in the following.   For the case $Q=0.8, b=0.8$, we find that the extremal black hole  mass is $M_{e}=0.7671287$. The variation of entropy and horizon for different masses are given in Table 4. For $k=0.01$, we  find that the variation of the entropy is negative for the case $M\leq 0.8$  and positive for  $M>0.8$. As $k$ increases to 0.05, the dividing point jumps to 1.35. From Figure (\ref{fig5}), in which the red curve and black curve represent cases for $M=0.8, 0.85$, we find that the black holes are single   horizon  black holes  for  $M=0.8, 0.85$. So we can also conclude that all the double horizon black holes  and part of the single horizon  black holes violate the second law of black hole thermodynamics.

\subsection{Weak cosmic censorship conjecture in the extended phase space}

In the extended phase space with consideration of pressure,  volume and Born-Infeld vacuum polarization, we  find that the first law of black hole thermodynamics is valid for both the single horizon black holes and double horizon black holes,  but the second law is violated  for the double horizon black holes and  part of the single horizon black holes.  In this section, we are going  to check the
weak cosmic censorship conjecture  in the extended phase space for the  Born-Infeld-anti-de Sitter. We want to know whether the horizon of the black hole will shrink to the singularity. As in the normal phase space, we will concentrate  on only the double horizon black holes since the weak cosmic censorship conjecture  is valid always for the single horizon black holes under charged particle absorbtion.

 As a charged particle is absorbed by the black hole, the mass $M$, charge $Q$,  Born-Infeld parameter $b$, and AdS radius $l$ change into $(M+dM, Q+dQ, b+db, l+dl)$. Correspondingly, the locations of
the minimum value and event horizon will change into $r_{m}\rightarrow r_{m}+dr_{m}$, $r_{+}\rightarrow r_{+}+dr_{+}$. There is also a transformation for $f(r)$, which is labeled as $df_{m}$. At the new  minimum point, according to   Eq.(\ref{eqc0}), we can get
\be \label{eqc2}
df'_{m}=\frac{\partial f'_{m}}{\partial M}dM+\frac{\partial f'_{m}}{\partial Q}dQ+\frac{\partial f'_{m}}{\partial l}dl+\frac{\partial f'_{m}}{\partial r_{m}}dr_{m}+\frac{\partial f'_{m}}{\partial b}dr_{b}=0.
\ee
In addition, at this location, $f(r)$ would change into
\be \label{eqc1}
f|_{r=r_{m}+dr_{m}}=f_{m}+df_{m}=\delta+\left(\frac{\partial f_{m}}{\partial M}dM+\frac{\partial f_{m}}{\partial Q}dQ+\frac{\partial f_{m}}{\partial l}dl+\frac{\partial f_{m}}{\partial b}db\right),
\ee
where we have used $f'_{m}=0$ in Eq.(\ref{eqc1}). Our next step is to find the final result of Eq.(\ref{eqc1}). We will discuss the extremal black hole, for which the horizon is located at $r_m$, so that   Eq.(\ref{eq:dispersion03}) can be used.

Based on the condition
$f'_{m}=0$, we obtain the concrete form of  $dM$ as
\be \label{eqc3}
dM=\frac{\partial M}{\partial r_m}dr_m +\frac{\partial M}{\partial Q}dQ+\frac{\partial M}{\partial l}dl+\frac{\partial M}{\partial b} db.
\ee
Substituting Eq.(\ref{eqc3}) into Eq.(\ref{eq:dispersion03}), we obtain
\be\label{eqc4}
dr_{m}=\frac{-4 b^2db l^3 r_m^4 \left(\sqrt{\frac{Q^2}{b^2 r_m^4}+1}-1\right)+2 b \left(r_m \sqrt{\frac{Q^2}{b^2 r_m^4}+1} \left(3 dl r_m^3-l^3|P^r_+|\right)+dQ l^3 Q\right)+2 db l^3 Q^2}{b l r_m \left(2 b k l^2 \sqrt{\frac{Q^2}{b^2 r_m^4}+1}+4 b^2 l^2 r_m^2 \left(\sqrt{\frac{Q^2}{b^2 r_m^4}+1}-1\right)+9 r_m^2 \sqrt{\frac{Q^2}{b^2 r_m^4}+1}\right)}.
\ee
In addition, substituting Eq.(\ref{eqc3}) into (\ref{eqc2}), we find $dr_{m}=0$. So with Eq.(\ref{eqc4}), we obtain
 \be \label{eqc5}
 dl=\frac{l^3 \left(2 b^2 db r_m^4 \left(\sqrt{\frac{Q^2}{b^2 r_m^4}+1}-1\right)+b \left(|P^r_+| r_m \sqrt{\frac{Q^2}{b^2 r_m^4}+1}-dQ Q\right)-db Q^2\right)}{3 b r_m^4 \sqrt{\frac{Q^2}{b^2 r_m^4}+1}}.
 \ee
 Similarly,
substituting Eq.(\ref{eqc3}) into Eq.(\ref{eqc1}), $f$ can be expressed as a function with respect to $dQ, dl, dr_{m}, db$. With the obtained results in Eq.(\ref{eqc4}) and Eq.(\ref{eqc5}), Eq.(\ref{eqc1}) can be simplified lastly as
\be \label{eqc6}
f_{m}+df_{m}= -\frac{2 |P^r_+|}{r_{m}}.
\ee
Eq.(\ref{eqc6}) shows that as a charged particle is absorbed by the black hole, there is a shift of $f_m$ in the negative direction.   So there always exist  horizons to hide the
singularity of the space time. In particular,  the extremal black holes will change into non-extremal black holes. Our result is different from the result in \cite{Gwak:2017kkt}, where they found that the black holes are stable, and the extremal black holes are always extremal black holes in the extended phase space.

\section{Conclusions }
\label{sec5}

In the Born-Infeld-anti-de Sitter  black hole, we investigated the motion of a charged particle around the black hole and obtained the relation between the energy and momentum of the particle near the horizon.
Furthermore, we  investigated the  laws of thermodynamics, and weak cosmic censorship conjecture in the normal phase space and extended phase space. As expected, the first law as well as  second law of the thermodynamics hold, and the weak cosmic censorship conjecture does not violate in the normal phase space. This is consistent with the previous conclusion that the charged black hole can not be over-spinning  as the backreaction is considered, even for the extremal black holes.
 In the extended phase space, since the cosmological constant and   Born-Infeld parameters are not a variable in the action and equations of the motion, the dynamical effect is not easy to predict with the extensive quantities $Y, b$. As the charged particle is absorbed by the black hole, we obtained the variation of the horizon of the black holes and further the variation of entropy and volume. We found that $T dS-YdV-b dB=|P^r_+|$, therefore, the  relation between the energy and momentum  can be written as the first law of thermodynamics in the extended phase space.
 It has been already proven that the satisfaction of the first law of thermodynamics is a necessary condition to ensure the second law of thermodynamics under a particle absorption \cite{Gwak:2015fsa,Gwak:2016gwj} in the normal phase space. But the satisfaction of the first law does not mean the second law is satisfied. So we also discussed the second law in the extended phase space. We found that
  the second law was violated for the double horizon black holes and part of the single horizon black holes.
 The violation of the second law of thermodynamics can be related to the weak cosmic censorship conjecture which is related to the stability of the horizon. The stability can be shown from the change of the minimal value of the function $f(r)$ under the absorption.
 We found that the variation of the minimal value  of  $f(r)$ in the extended phase space was consistent with that in the normal phase space, which is $- {2 |P^r_+|}/{r_{m}}$. In particular, the extremal
  black hole will change into non-extremal black holes. Our results show that the singularity will always be hidden behind the horizons, which implies that the weak cosmic censorship conjecture is valid in the extended phase space. Our result is different from  that in \cite{Gwak:2017kkt}, where they found that the black holes were stable, and the extremal black holes would always be extremal black holes. But our result is consistent with that in the normal phase space, namely the extremal black holes will change into the non-extremal black holes \cite{Gwak:2015fsa,Gwak:2016gwj}.

For the extremal black holes in the normal phase space, we know that the variation of the entropy is infinite, which implies that the horizon is infinite too according to the entropy area relation. So the singularity will be always hidden, which is consistent with the result obtained by weak cosmic censorship conjecture. For the extremal black hole in the extended phase space, we know that the variation  of the entropy is negative, the horizons of the black holes thus will shrink. However, we know that the weak cosmic censorship conjecture holds in this case, the horizon thus cannot shrink to the singularity. It is interesting to see that there exists the minimum value of the horizon where the black hole stops to shrink.
\section*{Acknowledgements}{Xiao-Xiong Zeng would like to thank
Bogeun Gwak and Hongbao Zhang for their helpful discussions. This work is supported  by the National
Natural Science Foundation of China (Grant Nos. 11875095, 11675140, 11705005), and Basic Research Project of Science and Technology Committee of Chongqing (Grant No. cstc2018jcyjA2480).}

%\bibliography{a}

\end{document}